\documentclass[aps,pra,twocolumn,showpacs,showkeys,superscriptaddress,reprint]{revtex4-1} 
\usepackage{graphics} 
\usepackage{graphicx} 
\usepackage{amsmath}
\usepackage{amssymb}
\usepackage{color}

\begin{document} 

\title{Insulator phases of a mixture of spinor fermions and hard-core bosons} 
\author{R. Avella}
\affiliation{Departamento de F\'{\i}sica, Universidad Nacional de Colombia, A. A. 5997 Bogot\'a, Colombia.}
\author{J. J. Mendoza-Arenas}
\affiliation{Departamento de F\'{\i}sica, Universidad de los Andes, A. A. 4976 Bogot\'a, Colombia.}
\author{R. Franco} 
\affiliation{Departamento de F\'{\i}sica, Universidad Nacional de Colombia, A. A. 5997 Bogot\'a, Colombia.}
\author{J. Silva-Valencia} 
\email{jsilvav@unal.edu.co} 
\affiliation{Departamento de F\'{\i}sica, Universidad Nacional de Colombia, A. A. 5997 Bogot\'a, Colombia.} 

\date{\today} 

\begin{abstract}
We study numerically a one-dimensional mixture of spin-$\tfrac{1}{2}$ fermions and scalar bosons in the hard-core limit. Considering repulsive fermion-fermion and boson-fermion interactions, we find superfluid and insulator states whose phase diagram is calculated. We determine that given a fermionic density $\rho_F$, the insulator states are located at the bosonic densities $\rho_B=1-\rho_F$ and $\rho_B=1-\tfrac{1}{2}\rho_F$, and emerge even in the absence of fermion-fermion coupling. In addition, the boson-fermion repulsion drives quantum phase transitions inside the insulator lobes with $\rho_B=1/2$. Our predictions could be observed in current cold-atom experimental platforms.
\end{abstract} 


\maketitle 

\section{\label{sec1}Introduction}
Mixtures of carriers obeying different statistics have intrigued scientists for decades, being the $^3$He-$^4$He system one of the first examples~\cite{Andreev-JETP75,Volovik-JETP75}. Now experimentalists can mix isotopes of the same atom, as well as different atoms, in clean and fully controllable set-ups, where the inter- and intra-species interactions can be tuned~\cite{Truscott-S01,Schreck-PRL01,Hadzibabic-PRL02,Roati-PRL02,Ott-PRL04,Silber-PRL05,Gunter-PRL06,Ospelkaus-PRL06,Zaccanti-PRA06,McNamara-PRL06,Best-PRL09,Fukuhara-PRA09b,YTakasu-JPSJ09,Deh-PRA10,Tey-PRA10,Sugawa-NP11,Schuster-PRA12,Tung-PRA13,Ferrier-Barbut-S14,Delehaye-PRL15,Vaidya-PRA15,XCYao-PRL16,YPWu-JPB17,Roy-PRL17,Schafer-PRA18}. These cold-atom setups have allowed researchers to observe a new mechanism of superfluid instability, phase separation, and a Bose-Fermi superfluid mixture~\cite{Trautmann-PRL18}, among other phenomena.\par 
From the theoretical point of view, the simplest approximation to describe a mixture of bosons and fermions is to freeze their internal degrees of freedom. This leads to a Bose-Fermi Hubbard Hamiltonian with free polarized fermions that interact with spinless bosons, which has been analyzed intensely by diverse analytical and numerical methods~\cite{Albus-PRA03,Cazalilla-PRL03,Lewenstein-PRL04,Mathey-PRL04,Roth-PRA04,Frahm-PRA05,Batchelor-PRA05,Takeuchi-PRA05,Pollet-PRL06,Mathey-PRA07,Mering-PRA08,Suzuki-PRA08,Luhmann-PRL08,Rizzi-PRA08,Orth-PRA09,XYin-PRA09,Sinha-PRB09,Orignac-PRA10,Polak-PRA10,Mering-PRA10,Masaki-JPSJ13}. From the above studies, a great variety of phases has been reported, including Mott insulators, different superfluids, spin and charge density waves, phase separation, and Wigner crystals. A broad study varying the bosonic density for a fixed density of polarized fermions was done by Zujev {\it et al.}~\cite{Zujev-PRA08}. Notably, using a quantum Monte Carlo method to explore a one-dimensional system, they found incompressible phases at the bosonic densities $\rho_B =1$ and $\rho_B=1-\rho_F$, being $\rho_F$ the fixed fermionic density. Note that the relation $\rho_B+\rho_F=1$ implies that the summation of the number of fermions and the number of bosons matches the system size, being this the so-called mixed Mott state of the mixture~\cite{Sugawa-NP11}.\par 
Motivated by the challenging and elusive physics of high-T$_c$ superconductors, several authors have investigated mixtures of unpolarized fermions and spinless bosons in two and three dimensions, making mean-field analytical or numerical calculations and taking into account in most cases half filling for fermions and one boson per site. Their results include $s$-, $p$- and $d$-wave pairing superconductivity, a charge density-wave phase (CDW), antiferromagnetic ordering, supersolid phases and superfluid-insulator transitions~\cite{Mathey-PRL06,Sengupta-PRA07,Mathey-PRB07,Klironomos-PRL07,Anders-PRL12,Bukov-PRB14,TOzawa-PRA14,Bilitewski-PRB15}.\par 
However the 1D case, in which low dimensionality is expected to result in diverse and new physics, has been barely explored. Using bosonization, Mathey {\it et al.} ~\cite{Mathey-PRL04} drew a first phase diagram of the model for non-commensurate fermionic densities and bosons lighter than fermions, showing CDW, spin density wave, phase separation, and single and triplet polaron pairing phases. This initial study motivates a numerical exploration of the rich physics of this model. In this paper, we study the ground state of a mixture of free hard-core bosons and spin-$\tfrac{1}{2}$ fermions confined in a 1D optical lattice, sweeping a wide range of values of the fermionic and bosonic densities. We go beyond mean field by performing calculations with the density matrix renormalization group technique~\cite{White-PRL92,Hallberg-AP06}, which allows us to fully consider correlations. For repulsive fermion-fermion and boson-fermion interactions, we obtain superfluid and insulator phases, and the transitions between them are explored. Notably, we show that given a fermionic density $\rho_F$, two insulator states arise at the bosonic densities $\rho_B=1-\rho_F$ and $\rho_B=1-\tfrac{1}{2}\rho_F$. Also we observe quantum phase transitions between insulator states only for a half-filling bosonic density.\par 
The paper is organized as follows. Our Bose-Fermi Hamiltonian is explained in Sec.~\ref{sec2}. The main results and phase diagrams for fermionic half-filling  and other densities are discussed in Secs.~\ref{sec3} and ~\ref{sec4}, respectively. In Sec.~\ref{sec5} we summarize our conclusions.\par 
\section{\label{sec2} Model}
We start by describing the model under consideration. The study of a mixture of bosonic and fermionic atoms is determined by the interaction term considered between bosons and fermions, which in our case will be the local repulsion between the atoms, leading to the Hamiltonian 
\begin{equation}\label{BFHamil}
    \hat{H}_{BF}=\hat{H}_B+\hat{H}_F+U_{BF}\sum_{i,\sigma}\hat{n}^{B}_{i}\hat{n}^{F}_{i,\sigma},
\end{equation}
with
\begin{equation}\label{FHamil}
    \hat{H}_{F}=-t_F\sum_{\langle i,j\rangle\sigma}\left(\hat{f}_{i,\sigma}^{\dag}\hat{f}_{j,\sigma} + \text{h.c.}\right)+U_{FF}\sum_{i}\hat{n}^{F}_{i,\uparrow}\hat{n}^{F}_{i,\downarrow},
\end{equation}
and
\begin{equation}\label{BHamil}
    \hat{H}_{B}=-t_B\sum_{\langle i,j\rangle}\left(\hat{b}_{i}^{\dag}\hat{b}_{j} + \text{h.c.}\right).
\end{equation}
The operator $\hat{b}_{i}^{\dag}$ ($\hat{f}_{i,\sigma}^{\dag}$) creates a boson (fermion of spin $\sigma=\uparrow,\downarrow$) at site $i$ of a lattice of size $L$. The local number operator for each species of fermions is $\hat{n}^{F}_{i,\sigma}=\hat{f}_{i,\sigma}^{\dag}\hat{f}_{i,\sigma}$, while the local boson number operator is $\hat{n}^{B}_{i}=\hat{b}_i^{\dag}\hat{b}_i$. The hopping matrix element between nearest-neighbor sites for bosons (fermions) is $t_B$ ($t_F$). The particles interact locally, where $U_{FF}$ and $U_{BF}$ denote fermion-fermion and boson-fermion interaction, respectively. We consider the hard-core limit for bosons, i.e. each site can be occupied by one boson at most, therefore there is no interaction term between bosons in Eq.~\eqref{BHamil}. Hence, the number of bosons per site (density of bosons $\rho_B$) takes values in the interval $[0,1]$. As usual the density for systems with two-color fermions varies in the interval $[0,2]$, such that $\rho_F=1$ corresponds to half-filling. We fix our energy scale taking $t_F=1$  in Eq.~\eqref{FHamil}.\par
Mixtures of degenerated gases of fermionic and bosonic atoms have been obtained in cold atoms setups, namely $^{6}$Li-$^{7}$Li \cite{Truscott-S01,Schreck-PRL01}, $^{40}$K-$^{87}$Rb \cite{Roati-PRL02} among others, although their stability is severely limited by 3-body recombinations. We highlight the mixtures obtained by the Takahashi's group \cite{YTakasu-JPSJ09}, $^{174}$Yb-$^{173}$Yb and $^{174}$Yb-$^{171}$Yb, where the $^{174}$Yb is a bosonic isotope with zero nuclear spin, while $^{173}$Yb and $^{171}$Yb are fermionic isotopes with nuclear spin $I=3/2$, and $1/2$, respectively. The above mixtures constitute promising candidates where our model Eq.~\eqref{BFHamil} could be studied in current cold-atom setups.\par 
\begin{figure}[t] 
\includegraphics[width=18pc]{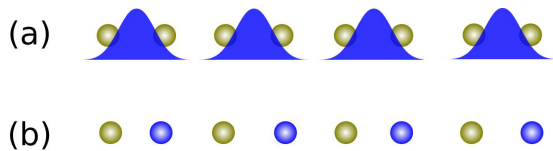} 
\caption{\label{fig0} Illustration of schematic ground states of a mixture of two-color fermions and hard-core bosons in one-dimension. Here, we consider a lattice with eight ($L=8$) sites, and two different ground states are depicted. (a) Insulator state with a fermionic density $\rho_F = 1$ and bosonic density $\rho_B = 1/2$. (b) Mixed Mott insulator state with $\rho_F = 1/2$, and $\rho_B = 1/2$.}
\end{figure} 
The ground state of our Bose-Fermi Hamiltonian ~\eqref{BFHamil} will be explored with the density matrix renormalization group algorithm (DMRG), using open boundary conditions. To perform the truncation, we use the dynamical block selection state (DBSS) protocol~\cite{Legeza-PRB03} keeping a discarded weight of $\sim10^{-7}$, and do finite-system sweeps until the ground-state energy is converged to an absolute error of $10^{-3}$.  In the following, $E(N_{\uparrow},N_{\downarrow},N_B)$ denotes the ground-state energy for $N_B$ bosons, and $N_{\uparrow}$, $N_{\downarrow}$ fermions.\par

\section{\label{sec3}  Half-filling case $\rho_F = 1$} 
\begin{figure}[t] 
\includegraphics[width=18pc]{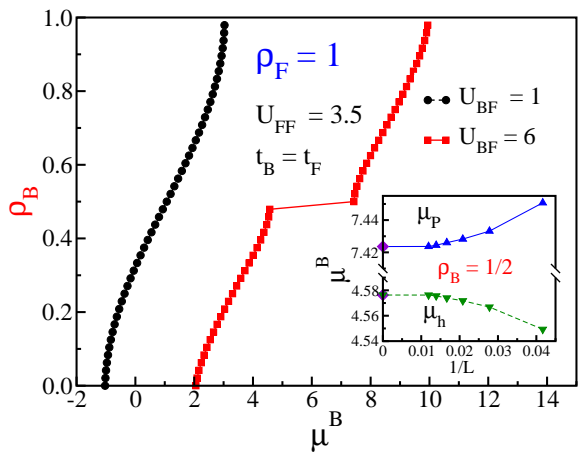} 
\caption{\label{fig1} Bosonic density $\rho_B$ as a function of chemical potential $\mu^B$. The fermion density is $\rho_F = 1$ (half-filling) and the fixed fermion-fermion repulsion is $U_{FF}=3.5$. Two different values of the boson-fermion interaction were considered, namely $U_{BF}=1$ (black) and $U_{BF}=6$ (red), for which incompressible phases are absent or present respectively. Inset: System size dependence of the chemical potential at the bosonic density $\rho_B = 1/2$. The upper set of data (blue) corresponds to the particle excitation energy and the lower one (green) to the hole excitation energy. The extrapolated diamond points, obtained by using a second-order polynomial function, indicate that in the thermodynamic limit there is a finite charge gap. The lines are visual guides.} 
\end{figure}

In the absence of bosons ($\rho_B=0$), we converge in the Fermi-Hubbard Hamiltonian, which at half-filling exhibits a Mott insulator state with antiferromagnetic order for any nonzero repulsive interaction. Adding bosons to the system makes the physics far more involved and rich, leading to a wide variety of  ground states.  One of these is depicted in Fig.~\ref{fig0} (a) where the bosonic density is $\rho_B = 1/2$ and the total number of carriers (bosons  plus fermions) is not commensurate with the lattice size. To explore the ground state of the mixture, we fix $U_{FF}=3.5$ and $U_{BF}=1$, obtaining that
 the chemical potential $\mu^B=E(N_{\uparrow},N_{\downarrow},N_B +1)-E(N_{\uparrow},N_{\downarrow},N_B)$ increases monotonously with the density, as can be seen in Fig.~\ref{fig1} (black dots). This indicates that it is not necessary to spend energy to generate bosonic excitations. However, for a larger boson-fermion repulsion $U_{BF}=6$ (red squares), the chemical potential initially increases monotonously with the number of bosons until a plateau emerges at the bosonic density $\rho_B=1/2$, after which the monotonic growth continues. The plateau  indicates that the ground state has a finite gap for this bosonic density. The energies for adding bosons $\mu_p=E(N_{\uparrow},N_{\downarrow},N_B +1)-E(N_{\uparrow},N_{\downarrow},N_B)$ and removing bosons $\mu_h=E(N_{\uparrow},N_{\downarrow},N_B)-E(N_{\uparrow},N_{\downarrow},N_B-1)$ as a function of $1/L$ are shown in the inset of Fig.~\ref{fig1}. We see that the above quantities remain unequal in the thermodynamic limit, giving a finite gap $\Delta^B=2.85$, which is the width of the plateau. For densities away from $\rho_B=1/2$ the energy for adding and removing bosons meet at the thermodynamic limit, and the ground state is superfluid.\par
The previous result is revealing because  it states that there is only one insulator state that fulfills the condition $\rho_B+\tfrac{1}{2}\rho_F=1$, for which the total number of carriers is non-commensurate with the lattice size. This is a notable finding since in the different approaches and experiments with Bose-Fermi mixtures the reported non-trivial insulators emerge due to the commensurability. Figure ~\ref{fig1} also suggests that for a given  fermion-fermion repulsion there is a critical value of the boson-fermion repulsion from which the insulator state appears at the bosonic density $\rho_B=1/2$.\par 
\begin{figure}[t!]
\begin{minipage}{19pc}
\includegraphics[width=19pc]{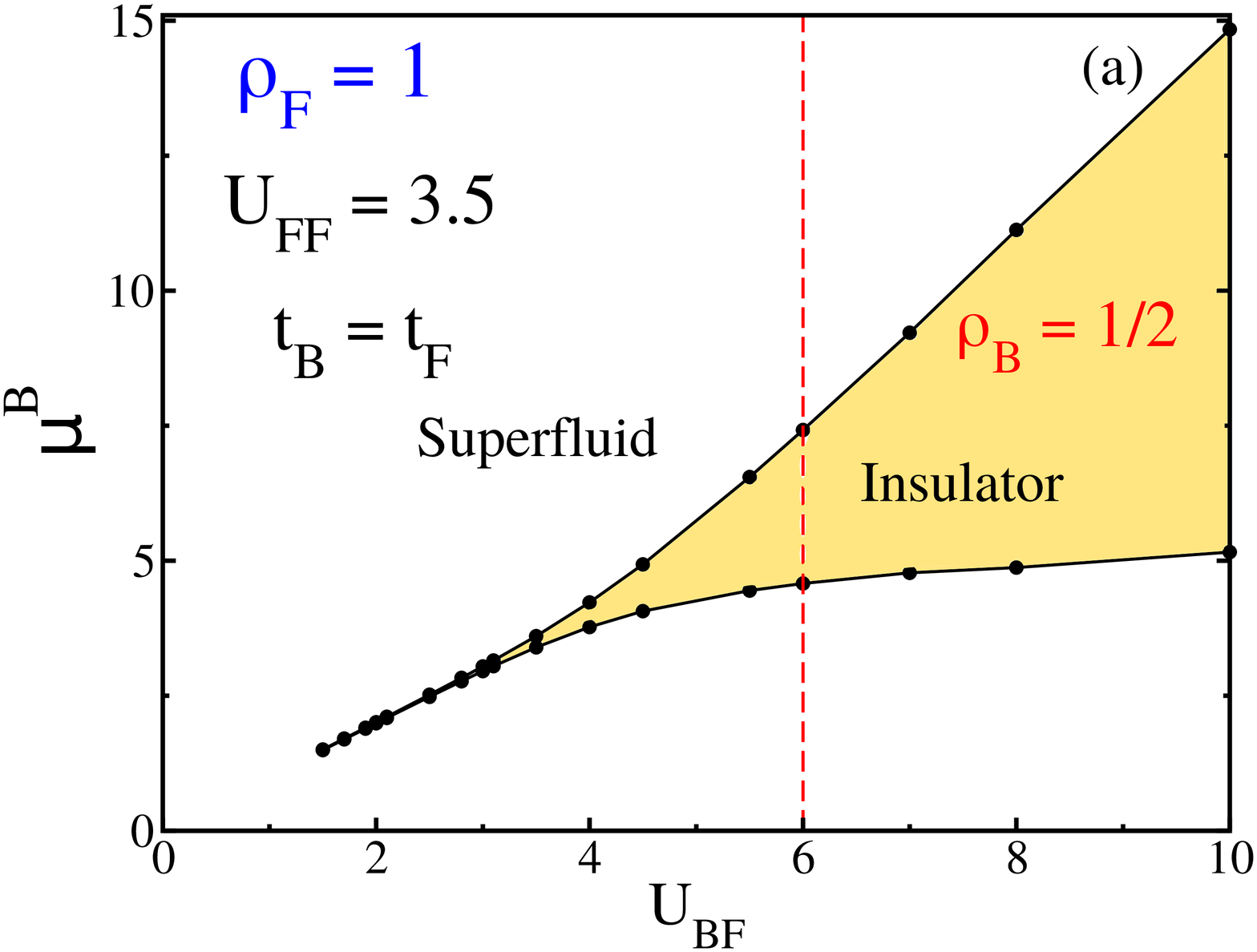}
\end{minipage}
\hspace{5pc}%
\begin{minipage}{19pc}
\includegraphics[width=19pc]{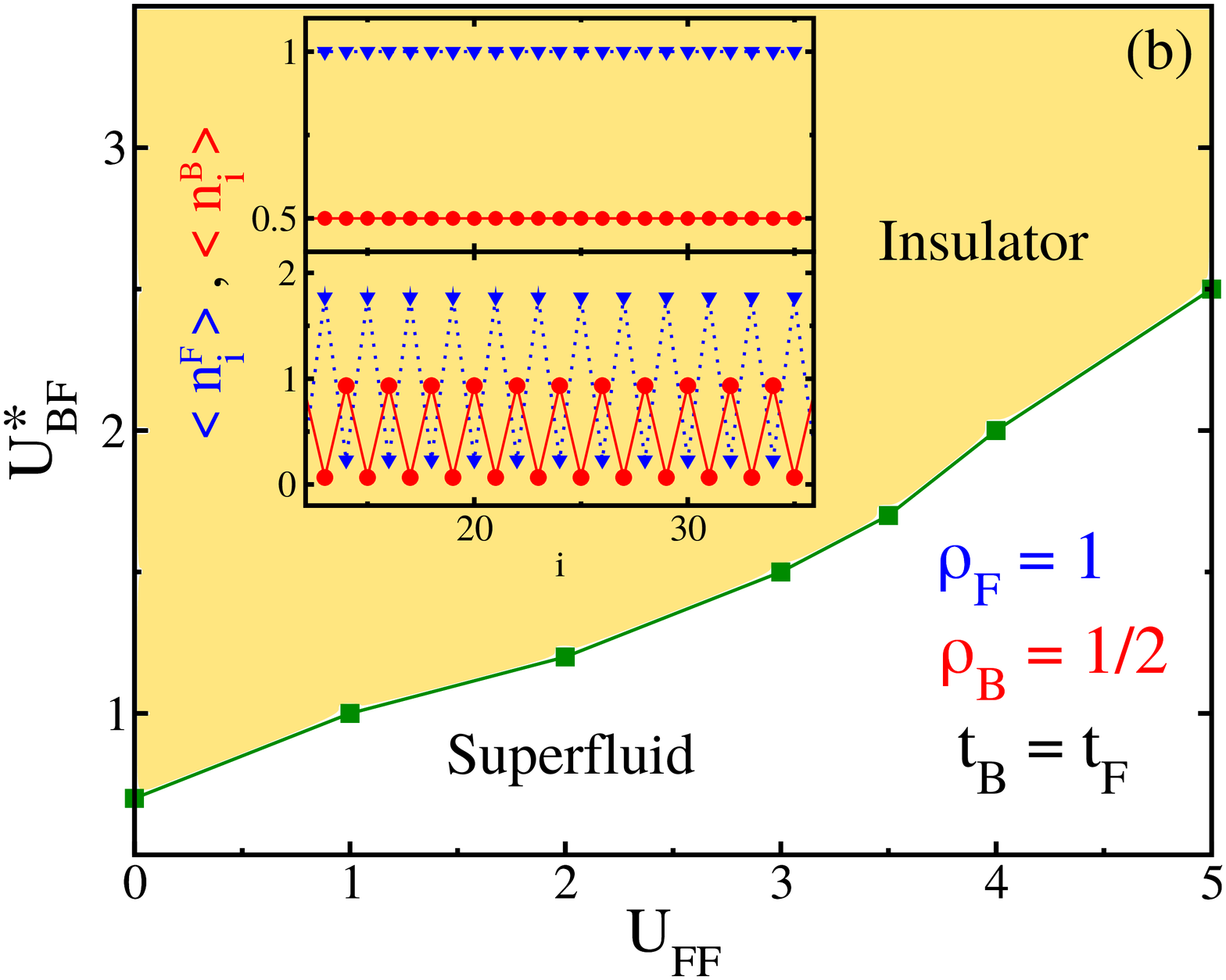}
\caption{\label{fig2} (a) Phase diagram showing the chemical potential versus the boson-fermion repulsion. The fermion density is $\rho_F = 1$ and the fixed fermion-fermion repulsion is $U_{FF}=3.5$. The superfluid phase is shown in white, while the yellow zone corresponds to an insulator phase for $\rho_B = 1/2$, which appears from a critical interaction value $U^{*}_{BF}$. The vertical dashed line corresponds to the curve of $\rho_B$  vs chemical potencial shown in Fig.~\ref{fig1} (red squares) for $U_{BF}=6$. (b) Critical points $U^{*}_{BF}$ as a function of $U_{FF}$. The incompressible phase fulfills $\rho_B+\tfrac{1}{2}\rho_F=1$. In the inset, we show the density profile for bosons (red circles) and fermions (blue triangles) for $U_{BF}=5$ (top panel) and  $U_{BF}=9$ (bottom panel), with $U_{FF}=3.5$. The lines are visual guides.}
\end{minipage}\hspace{2pc}%
\end{figure}
Replicating Fig.~\ref{fig1} for others values of $U_{BF}$, we obtained the phase diagram at fixed $U_{FF}=3.5$ in the $\mu^B$-$U_{BF}$ plane shown in Fig.~\ref{fig2} (a). In the absence of repulsion between bosons and fermions it costs no energy in the thermodynamic limit to generate bosonic excitations. This means that the chemical potential to add or remove a boson is the same and the ground state is a superfluid. Such a behavior  remains up to a critical value $U^{*}_{BF}\approx1.50$, from which the chemical potentials differ, the plateau at $\rho_B=1/2$ emerges and grows as a function of the boson-fermion repulsion (yellow area). Similar (not shown) phase diagrams with the insulator phase at $\rho_B=1/2$ surrounded by superfluid phases were obtained for others values of the fermion-fermion repulsion.\par
The critical value $U^{*}_{BF}$ for which the insulator state emerges at $\rho_B=1/2$ varies with the fermion-fermion repulsion; its evolution is shown in Fig.~\ref{fig2} (b). First we note a remarkable result, namely that even with no fermion-fermion repulsion, there is a critical value of the boson-fermion repulsion from which the insulator state at $\rho_B=1/2$ appears. This can be intuitively explained since in such a limit, a strong enough boson-fermion local interaction can create stable heavy quasiparticles that do not decay in spite of its repulsive nature. This occurs because the associated potential energy is larger than the maximal kinetic energy for independent particles, so the former cannot be converted into the latter. These massive repulsively bound pairs propagate very slowly across the lattice, resulting in an insulating state~\cite{Winkler-Nat06}. To the best of our knowledge, this result has not been reported before in any boson-fermion mixture system. When turning on the fermion-fermion repulsion, we see that the critical value grows with $U_{FF}$, dividing the phase diagram into a gapless (lower part) and a gapfull phase (upper part) associated with the insulator defined by the condition $\rho_B+\tfrac{1}{2}\rho_F=1$. This  trend emerges because when increasing $U_{FF}$, fermions tend to move away from each other even more, reducing the impact on the bosons. Thus to keep the bound states and thus the homogeneous insulating phase (see the upper panel of the inset of Fig.~\ref{fig2} (b)), it is necessary to increase the boson-fermion coupling.\par 
Now we discuss how bosons and fermions are distributed in the insulator state. For weak boson-fermion interactions, we found one fermion per site and one boson extended across two sites, as depicted in the top panel of the inset of Fig.~\ref{fig2} (b) for $U_{BF}=5$ and sketched in Fig.~\ref{fig0} (a). The above result suggests that the fermion sector provides a Mott background under which a superfluid-gapped state transition takes place as the number of bosons vary, consistent with the description of the previous paragraph. Note that, considering global commensurate densities for both bosons and fermions, a superfluid-Mott insulator transition under a fermionic CDW background was reported for a Bose-Fermi mixture in 3D~\cite{Anders-PRL12,Bukov-PRB14}. For larger boson-fermion repulsion the fermions will not occupy the same sites as the bosons, so the distribution will not be homogeneous. This is exemplified in the bottom panel of the inset of Fig.~\ref{fig2} (b), where we consider $U_{BF}=9$. There we obtain an interleaved band-insulator state, which has a two-site unit cell with one boson in a site and a local doublon of fermions in the other. Thus we conclude that within the insulator lobe, a quantum phase transition between the states shown in the inset of Fig.~\ref{fig2} (b) takes place for each value of the fermion-fermion interaction, whose critical points will be reported in a future work. Note that similar configurations, namely CDW phases, and spin density wave to CDW transitions, have been predicted by the bosonization approach~\cite{Mathey-PRL04}. Also, a similar ground state characterized by a two-site unit cell with bosons and fermions in different sites has been reported for polarized fermions and spinless bosons mixture with a density of half-filling for both kind of carriers~\cite{Titvinidze-PRL08}.\par

\section{\label{sec4} Other fillings}
\begin{figure}[t] 
\includegraphics[width=18pc]{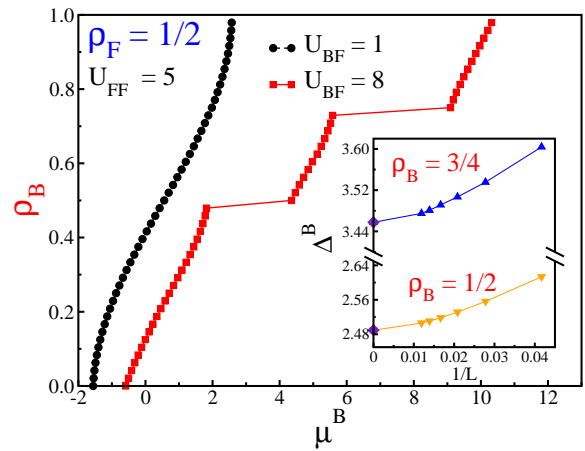} 
\caption{\label{fig3} Bosonic density $\rho_B$ as a function of chemical potential $\mu_B$ for fixed fermionic density $\rho_F = 1/2$ and  fermion-fermion repulsion $U_{FF}=5$. Two different values of the boson-fermion interaction were considered, $U_{BF}=1$ (black dots) and $U_{BF}=8$ (red squares), for which incomprenssible phases are absent or present respectively. Inset: System size dependence of the bosonic charge gap  at the densities $\rho_B = 1/2$ (orange) and $\rho_B = 3/4$ (blue). The values for $1/L\rightarrow0$ (diamonds) correspond to a second-order polynomial extrapolation to the thermodynamic limit. The lines are visual guides.} 
\end{figure}

The above results reveal an insulator state with a fermionic density commensurate with the lattice size; however it is important to move beyond this condition and explore other fillings. Initially we consider fermionic quarter filling, $\rho_F = 1/2$. In Fig.~\ref{fig3}, we display the bosonic density as a function of the chemical potential at fixed fermion-fermion repulsion $U_{FF}=5$. For a weak boson-fermion interaction $U_{BF}=1$, we observe that the chemical potential evolves continuously and no plateaus arise, hence the ground state is a superfluid. But novel features are found for larger values of the boson-fermion repulsion, namely two plateaus arise at the bosonic densities $\rho_B=1/2$ and $\rho_B=3/4$ indicating insulator states at these densities. In the inset of Fig.~\ref{fig3}, we show the evolution of the width of these plateaus as a function of the inverse of the lattice size, indicating that in the thermodynamic limit they reach a finite value $\Delta^B=2.49$ and $\Delta^B=3.46$ for $\rho_B=1/2$ and $\rho_B=3/4$, respectively.\par
Considering that an insulator state arises for the bosonic density $\rho_B=1/2$ and the fact that the fixed fermionic density is $\rho_F = 1/2$, we see that the relation $\rho_F+\rho_B=1$ is satisfied, showing that the total number of carriers is commensurable with the lattice size. The relation $\rho_F+\rho_B=1$, with the correct interpretation of $\rho_F$, has been obtained for insulating states of Bose-Fermi mixtures in different contexts, for instance in mixtures of bosons with polarized~\cite{Zujev-PRA08} and spinful~\cite{Sugawa-NP11} fermions. In this mixed Mott state, the total number of carriers is equal to the lattice size, even though the number of fermions or bosons are not commensurate (see Fig.~\ref{fig0} (b)).\par 
The widest plateau in Fig.~\ref{fig3} corresponds to the bosonic density  $\rho_B=3/4$ and the relation $\rho_B+\tfrac{1}{2}\rho_F=1$ is maintained, which is related to the new insulating state reported in the previous section. Therefore, we conclude that the above relation always determines a plateau.\par
For other fermionic densities, such as $\rho_F = 1/3$ we verified that the above relations between fermionic and bosonic densties are fulfilled, which is a main conclusion of the present study.\par
\begin{figure}[t!]
\begin{minipage}{19pc}
\includegraphics[width=19pc]{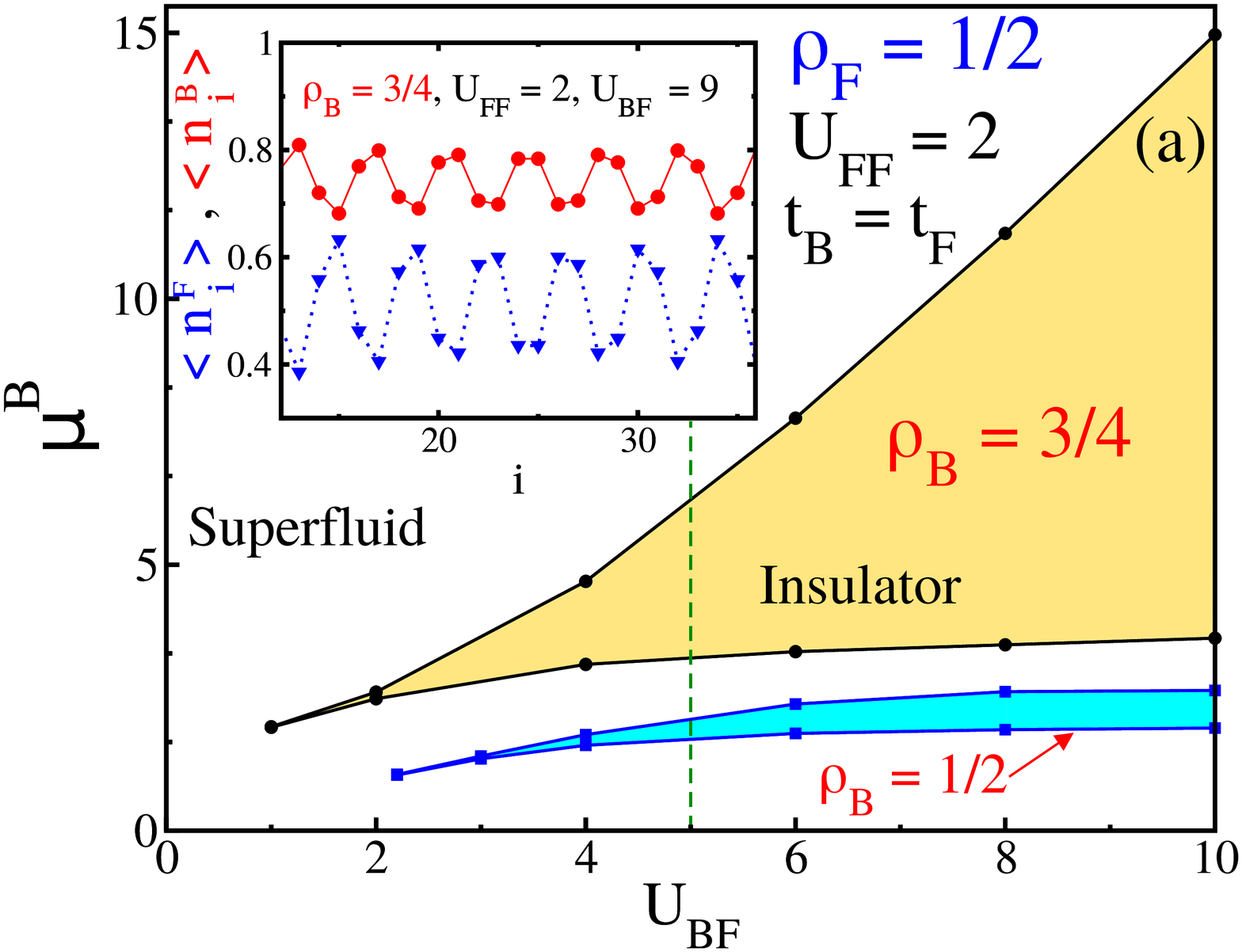}
\end{minipage}
\hspace{5pc}%
\begin{minipage}{19pc}
\includegraphics[width=19pc]{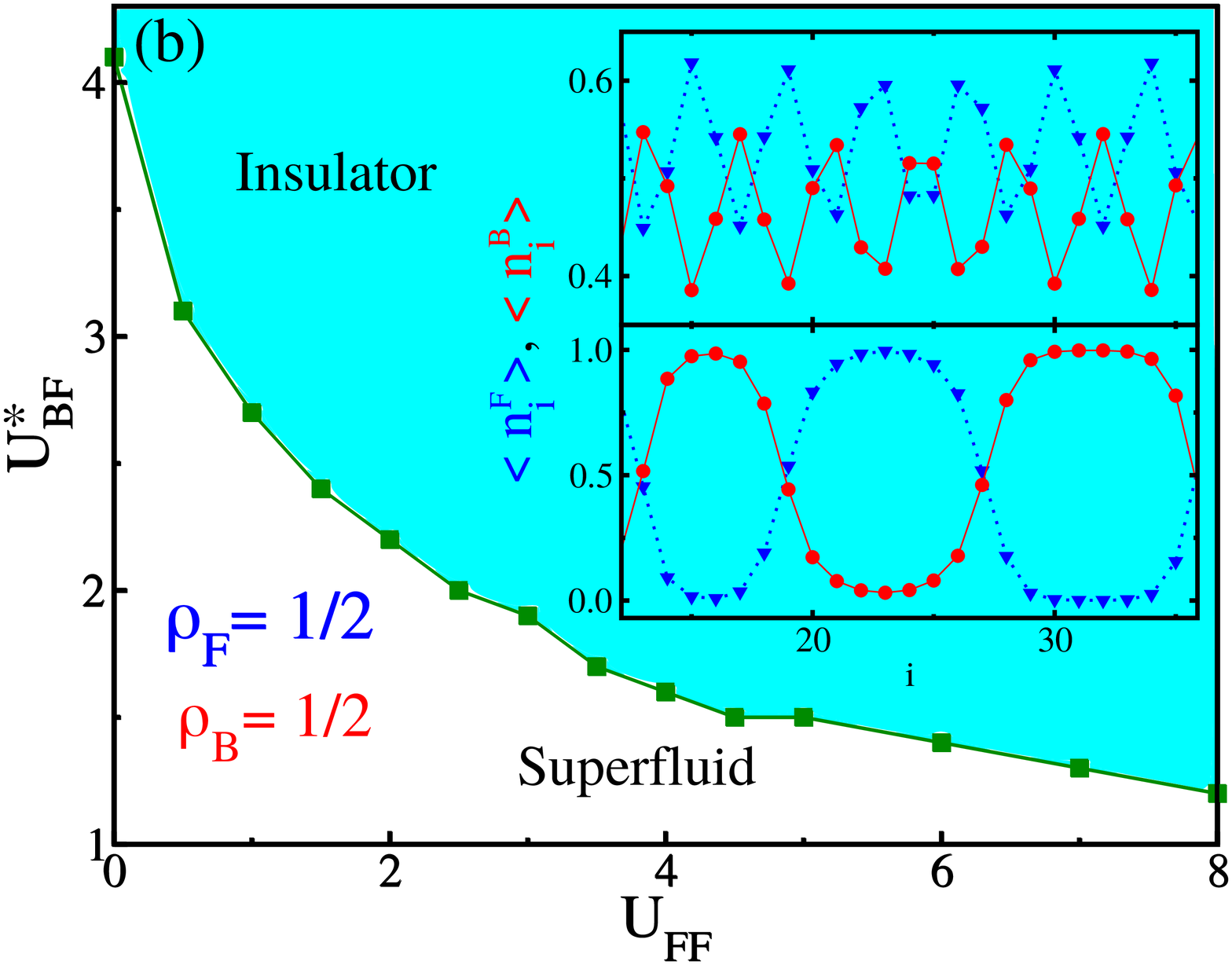}
\caption{\label{fig4} (a) Ground state phase diagram in the $\mu^B$-$U_{BF}$ plane for fixed  fermion-fermion repulsion $U_{FF}=2$ and fermionic  density $\rho_F = 1/2$. The colored regions correspond to incompressible states. The yellow region is for $\rho_B=3/4$, while the blue one is for $\rho_B=1/2$. The distribution of fermions (blue triangles) and bosons (red circles) across the lattice is shown in the inset; the ground state within the yellow lobe ($\rho_B=3/4$) is a CDW regardless the boson-fermion interaction. (b) Critical points as a function of the fermion-fermion interaction for the density $\rho_B=1/2$. The points separate a gapfull phase from a superfluid one. In the inset, we show that within the blue lobe different density profiles are obtained for $U_{BF}=13$ (top panel) and  $U_{BF}=16$ (bottom panel), considering $U_{FF}=2$. This corresponds to is a quantum phase transition from a CDW to a phase-separation state. The lines are visual guides.}
\end{minipage}\hspace{2pc}%
\end{figure}

The emergence of two plateaus for a fermionic density $\rho_F = 1/2$ suggests a rich ground-state physics; we replicate Fig.~\ref{fig3} for others values of $U_{BF}$  and obtain the phase diagram depicted Fig.~\ref{fig4} (a), where the chemical potential in the thermodynamic limit is displayed for several values of  $U_{BF}$ keeping the fermion-fermion interaction $U_{FF}=2$. Colored regions indicate insulator states, namely the blue lobe corresponds to a mixed Mott state with $\rho_B=1/2$, while the yellow lobe is associated to the insulator state with $\rho_B=3/4$. The phase diagram suggests three different superfluid regions, which surround the insulator lobes. Notably, these lobes have different shapes. For bosonic density $\rho_B=1/2$ both the energy for adding and removing a boson increases slowly and their difference tends to remain constant for larger values of the boson-fermion interaction, while a much larger difference takes place for $\rho_B=3/4$. Comparing  Fig.~\ref{fig2} (a) and Fig.~\ref{fig4} (a), we can state that insulator lobes for which $\rho_B+\tfrac{1}{2}\rho_F=1$ are characterized by an energy for removing a boson that increases slowly, tending to saturate, while the energy for adding a boson increases rapidly with the boson-fermion interaction. We also found that similarly to Fig.~\ref{fig2} (b), the critical point $U^{*}_{BF}$ from which the insulator phase emerges increases with the fermion-fermion interaction (not shown for $\rho_B=3/4$). We conclude that the lobes that fulfill the relation $\rho_B+\tfrac{1}{2}\rho_F=1$ possess features not shared by the others.\par
In the inset of Fig.~\ref{fig4} (a), we show the density profile for bosons and fermions for $\rho_B=3/4$, $U_{FF}=2$, and $U_{BF}=9$. It is clear that the ground state is a two-site periodic modulation of charge, i.e. a charge density wave, which remains qualitatively the same regardless of the boson-fermion interaction.\par 
Given a fermion-fermion interaction, we observe that the insulator lobes arise from different values of the boson-fermion coupling (see Fig.~\ref{fig4} (a)), which is expected because for the same number of interacting fermions, there are more bosons in one lobe than in the other. The critical points for the insulator state with $\rho_B=1/2$ are shown in Fig.~\ref{fig4} (b) for different values of the fermion-fermion interaction. In the absence of the latter and with a few bosons in the lattice, a large value of $U^{*}_{BF}\approx 4.2$ is required to distribute and localize the carriers on different sites (as sketched in Fig.~\ref{fig0} (b))) and generate an insulating state. This is much larger than that required to induce the insulator at half filling and $\rho_B+\tfrac{1}{2}\rho_F=1$ (see Fig.~\ref{fig2} (b)), due to the higher amount of fermions in that case. When turning on fermionic repulsion, a lower boson-fermion interaction is enough to localize the particles. Above these critical points (green squares) the state is insulating, whereas below them it is superfluid. Note that the above evolution of criticality contrasts to the findings for the lobes that fulfill the relation $\rho_B+\tfrac{1}{2}\rho_F=1$, for which the critical points increase with $U_{FF}$ (see Fig.~\ref{fig2} (b)). This is the case because due to the lower fermionic density, which tends to locate bosons and fermions on different sites, a weaker boson-fermion coupling is needed to keep an insulating state when the fermion-fermion repulsion is finite.\par 
We note that the distribution of bosons and fermions in the insulator lobes is diverse and it can depend of the boson-fermion interaction. For the insulator lobe with bosonic density $\rho_B=1/2$, we display the density profile of bosons and fermions in the inset of Fig.~\ref{fig4} (b) for two different values of $U_{BF}$, keeping the fermion-fermion interaction fixed. In the top panel, we consider $U_{BF}=13$ and observe that the ground state is a charge density wave, but the density profiles are interwoven due to the low global density. Increasing the boson-fermion repulsion ($U_{BF}=16$) and due to the low number of carriers, regions with bunching of fermions and no bosons and vice versa emerge along the lattice establishing a phase-separation state, which has been previously reported~\cite{Mathey-PRL04,Mathey-PRA07,Sugawa-NP11} and recently observed in a Bose-Fermi mixture of 
$^{41}K$ and $^{6}$Li atoms~\cite{Lous-PRL18}. The density-density correlations for both bosons and fermions allow us to characterize the ground state. From this we obtain that the charge structure factor $\mathcal{N}^{F,B}(q)=\sum_{j,l=1}^{L}e^{iq(j-l)}\left(\langle n_{j}^{F,B} n_{l}^{F,B}\rangle-\langle n_{j}^{F,B}\rangle\langle n_{l}^{F,B}\rangle\right)$ for fermions and bosons exhibit a maximum around $q/\pi\sim0.54$ for $U_{BF}=13$, while a monotonic growth was observed for $U_{BF}=16$. Thus, inside the insulator lobe with a bosonic density $\rho_B=1/2$ a quantum phase transition between different states is driven by the boson-fermion repulsion.\par 
For fermionic densities larger than half-filling, a scenario similar to the one discussed above emerges, with the presence of some plateaus as the number of bosonic atoms increases. For instance, when the fermionic density is $\rho_F=3/2$ insulator lobes arise for bosonic densities $\rho_B=1/4$ and $\rho_B=1/2$ due to the fermionic particle-hole symmetry.\par
\begin{figure}[t] 
\includegraphics[width=18pc]{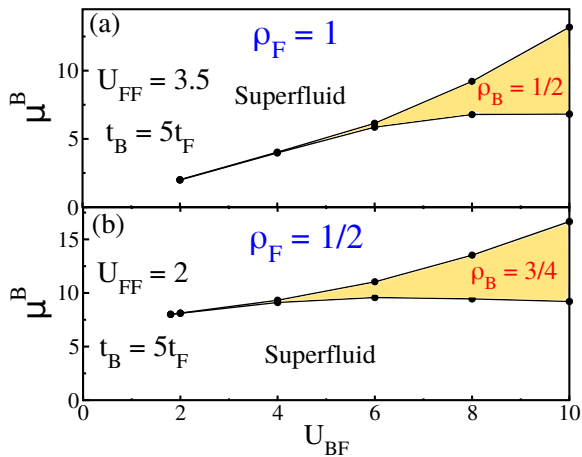} 
\caption{\label{fig5} Phase diagram showing the chemical potential versus the boson-fermion repulsion, for a mass imbalance $t_B= 5t_F$. (a) Fermionic density $\rho_F = 1$ and fixed fermion-fermion repulsion $U_{FF}=3.5$. (b) Fermionic density $\rho_F = 1/2$ and fixed fermion-fermion repulsion $U_{FF}=2$. The superfluid phase is shown in white, while the colored areas correspond to incompressible regions that fulfill $\rho_B+\tfrac{1}{2}\rho_F=1$. The lines are visual guides.} 
\end{figure} 
\section{\label{sec5} Mass imbalance}
Experimentalists in cold atom systems have been able to confine diverse bosonic and fermionic atoms simultaneously, which enriches the physics even further. Different atoms imply different masses, which can be modeled by taking different hopping parameters~\cite{Sowinski-RPP19}. We consider different values of $t_B/t_F$ and observe that for $t_B/t_F\leq 1$, the plateaus show in Figs.~\ref{fig1} and~\ref{fig3} survive. However surprises arise when we consider values greater than one. To exemplify them, the effect of a mass imbalance on the results of Secs.~\ref{sec3} and~\ref{sec4} was analized for $t_B/t_F=5$, which has been considered in previous studies about Bose-Fermi mixtures~\cite{Illuminati-PRL04,Mathey-PRA07}.\par 
In Fig.~\ref{fig5} (a), we consider the same parameters that in Fig.~\ref{fig2} (a), and see that the insulator lobe fulfilling $\rho_B+\tfrac{1}{2}\rho_F=1$ survives. However its area in the phase diagram becomes more narrow as expected due to the enhanced dynamics of the bosonic carriers. Also, we note that the critical boson-fermion coupling from which the insulator phase appears shifts to larger values. Similar results are obtained for a lower fermionic density $\rho_F=1/2$, but the second insulator region shown in Fig.~\ref{fig4} (a) disappears due to the mass imbalance.\par 
\section{\label{sec6} Conclusions}
The ground state of a mixture of spin-$\tfrac{1}{2}$ fermions and hard-core bosons described by the Bose-Fermi Hubbard Hamiltonian was explored using the density matrix renormalization group  technique. The repulsive fermion-fermion and boson-fermion interactions were varied, from which 
phase diagrams for different fermionic densities were built. For the fermionic half-filling case ($\rho_F = 1$), we found a unique insulator state as the number of bosons varies. This precisely occurs at $\rho_B=1/2$, leading to the relation $\rho_B+\tfrac{1}{2}\rho_F=1$. Within the insulator lobe a quantum phase transition driven by the boson-fermion repulsion takes place, where the ground state passes form a Mott state with one fermion per site and one boson shared by two sites to an interleaved band insulators state with a two-site unit cell with two fermions in one site and a boson in the other site.\par  
Two insulating states arise away from half-filling, which are located at the bosonic densities $\rho_B=1-\rho_F$ and $\rho_B=1-\tfrac{1}{2}\rho_F$, for a give fermionic density $\rho_F$. The former are different from the latter, namely the nature of the insulating ground state for fillings $\rho_F = 1/2$ and $\rho_B=3/4$ remains unaltered as the boson-fermion interaction increases, while for the lower bosonic density $\rho_B=1/2$ a quantum phase transition from a CDW state to a phase separation was observed.\par 
In addition, we found that the insulator lobes emerge for a finite critical value of the boson-fermion interaction, whose evolution with the fermion-fermion interaction depends on whether the lobe fulfills the relation $\rho_B+\tfrac{1}{2}\rho_F=1$ or not.\par
Finally, a mass imbalance between bosons and fermions leads to a narrowing of the lobes that fulfilled $\rho_B+\tfrac{1}{2}\rho_F=1$, while the others can disappear depending of the interaction parameters.\par
We believe that our findings can help interpret experimental results in atomic mixtures. In addition, the states and transitions reported here can be tested in current cold atom setups.\par 
%
\section*{Acknowledgments}
R. A. thanks the support of Departamento Administrativo de Ciencia, Tecnolog\'{\i}a e Innovaci\'on (COLCIENCIAS)
(Grant No. FP44842-135-2017). J. S.-V., R. F. and R. A. are thankful for the support of DIEB- Universidad Nacional de Colombia (Grant No. 42133).
J. J. M.-A. thanks the support of Departamento Administrativo de Ciencia, Tecnolog\'{\i}a e Innovaci\'on (COLCIENCIAS), through the project \textit{Producci\'on y Caracterizaci\'on de Nuevos Materiales Cu\'anticos de Baja Dimensionalidad: Criticalidad Cu\'antica y Transiciones de Fase Electr\'onicas} (Grant No. 120480863414). J. J. M.-A. thanks the Galileo Galilei Institute for Theoretical Physics for the hospitality and the INFN for partial support during the completion of this work.


%
\bibliography{Bibliografia}

\end{document}